\documentclass[a4paper,10pt]{article}
\usepackage{enumerate}
\usepackage{color}
\usepackage[utf8]{inputenc} 
\usepackage[english]{babel}
\usepackage[T1]{fontenc}
\usepackage{graphicx}
\usepackage{amsfonts,amssymb,amsmath,latexsym,amsthm}
\usepackage{textcomp}
\usepackage[pdftex]{hyperref}
\usepackage{geometry}
\title{Solitary wave interactions with a  periodic forcing: the extended Korteweg-de Vries framework}
\author{Marcelo V. Flamarion$^{1}$ and Efim Pelinovsky$^{2,3}$}
\date{}

\begin{document}
\maketitle
\begin{center}
{\footnotesize $^1$Unidade Acad{\^ e}mica do Cabo de Santo Agostinho, \\
UFRPE/Rural Federal University of Pernambuco, BR 101 Sul, Cabo de Santo Agostinho-PE, Brazil,  54503-900 \\
marcelo.flamarion@ufrpe.br }

\vspace{0.3cm}
{\footnotesize $^{2}$Institute of Applied Physics, 46 Uljanov Str., Nizhny Novgorod 603155, Russia. \\
 $^{3}$National Research University--Higher School of Economics, Moscow, Russia. }

%{\footnotesize ORCID Number: 0000-0002-0877-4831}

\end{center}

%\maketitle

\begin{abstract} 
The aim of this work is to study numerically the interaction of large amplitude solitary waves with an external periodic forcing using the forced extended Korteweg-de Vries equation (feKdV). Regarding these interactions,  we find that a solitary wave can bounce back and forth remaining close to its initial position when the forcing and the solitary wave are near resonant  or it can move away from its initial position without reversing their direction. Additionally, we verify that the numerical results agree well within the asymptotic approximation for broad the forcings. %when the topography varies rapidly and the solitary wave propagates with the same speed as the current in opposite direction, the solitary wave remain almost steady for large times. In other words, current and topography barely affect the solitary wave behaviour.

	\end{abstract}

\section{Introduction}
The forced extended Korteweg-de Vries equation (efKdV) is usually used  to describe internal solitary waves in ocean over a variable topography in the presence of a  current when the nonlinearity is stronger than the one predicted by the forced Korteweg-de Vries equation (fKdV). In recent studies a great attention has been paid to trapped waves \cite{Ermakov, Lee,LeeWhang,Kim, COAM, Collisions, Capillary2}, which are described as waves that  bounce back and forth above an obstacle remaining trapped for large times.

A complete asymptotic study on trapped waves for the fKdV equation was done by Grimshaw and collaborators \cite{Grimshaw94, Grimshaw96}
for a localized topography. In these works, the authors found that the asymptotic results agree with the numerical predictions. Along the same lines, Grimshaw and Pelinovsky \cite{Grimshaw2002} investigated asymptotically  trapped solitary waves for the efKdV equation.

Regarding periodic topographies, Malomed \cite{Malomed:1993} investigated asymptotically the emission of radiation of solitons in a periodic forcing for the fKdV equation. He showed that the forcing does not capture the solitons. In fact, under the action of the radiative losses, a soliton which was moving slower than the forcing is further decelerated, while the one which was faster is accelerated. Numerical results confirming his findings were reported later by Grimshaw et al. \cite{Grimshaw:1993}.

In this article, differently from the mentioned works above, we focus on trapped waves due to a periodic forcing for the efKdV equation.  We show that for a solitary wave with speed close to the forcing can be trapped for large times or it can propagate away from its initial position without reversing its direction. Besides, the numerical results are confirmed within the asymptotic framework for broad forcings.

This article is organized as follows. In section 2 we present the mathematical formulation of the problem. Numerical methods and results are presented in section 3 and  conclusion in section 4.

\section{The forced extended Korteweg-de Vries equation}
We consider the  extended Korteweg-de Vries equation with a forcing term as
model to study trapped waves 
\begin{equation}\label{efKdV1}
U_{t} +6UU_{x}+U^{2} U_x+U_{xxx}=\epsilon f_{x}(x+\Delta t).
\end{equation}
Here, we denote by $U(x,t)$ denotes  the surface profile, $f(x+\Delta t)$  the periodic forcing that travels with constant speed $\Delta$ and $\epsilon>0$ is a small parameter.  It is convenient to rewrite equation (\ref{efKdV}) in the forcing moving frame. Therefore, 
\begin{equation}\label{efKdV}
U_{t} +\Delta U_x+ 6UU_{x}+U^{2} U_x+U_{xxx}=\epsilon f_{x}(x).
\end{equation}
This equation conserves mass $(M(t))$, with
\begin{equation}\label{mass}
\frac{dM}{dt}=0, \mbox {where } M(t)=\int_{-\infty}^{\infty}U(x,t)dx,
\end{equation} 
and the rate of change of momentum $(P(t))$ is balanced by the external forcing as
\begin{equation}\label{momentum}
\frac{dP}{dt}=\int_{-\infty}^{\infty}U(x,t)\frac{df(x)}{dx}dx, \mbox {where } P(t)=\frac{1}{2}\int_{-\infty}^{\infty}U^{2}(x,t)dx.
\end{equation} 
In the absence of an external forcing, the eKdV admits two families of solitary waves as solutions  \cite{Grimshaw:2010}, which are given by the expressions
\begin{equation}\label{solitary}
U(x,t)=\frac{\gamma^2}{1+B\cosh(\gamma(x-ct))}, \mbox{  where }  \;\ c=\Delta+\gamma^2, \;\ B^{2} = 1 +\frac{ \gamma^2}{6}.
\end{equation}
Here we analyze only elevation solitary waves ($B>0$) whose amplitude  is 
\begin{equation} \label{amplitude}
a=\frac{\gamma^{2}}{1+B}=6(B-1).
\end{equation}
The periodic forcing is modeled by the function 
\begin{equation}\label{forcing}
f(x) = A\sin(qx), 
\end{equation}
where $A$ is its amplitude and $q$ is the wave number. Since the perturbation $f$ is not localized it produces radiation all over the domain, even far away from where most of the energy of the  solitary wave is localized. Although it does not affect the asymptotic study at lowest orders, it can be troublesome for the numerical study. For this reason, we use a similar trick as done by Malomed \cite{Malomed:1993}. Inserting into equation (\ref{efKdV}) $$U(x,t)=u(x,t)+\epsilon u_{0}(x),$$ 
where $$u_{0}(x)=\frac{A}{\Delta-q^2}\sin(qx)$$
is the solution of the linearized feKdV equation (\ref{efKdV}), we have that $u(x,t)$ satisfies 
$$u_{t}+6 uu_x+ u^{2}u_x +u_{xxx}=-6\epsilon(u_{0}u)_{x}-\epsilon(u^{2}u_{0})_x+\mathcal{O}(\epsilon^{2}).$$
Consequently, at first approximation we obtain the new equation
\begin{equation}\label{UefKdV}
u_{t}+6 uu_x+u^{2}u_x +u_{xxx}=-6\epsilon(u_{0}u)_{x}-\epsilon(u^{2}u_{0})_x,
\end{equation}
where the perturbation now is localized along the free surface $u(x,t)$.

%There are a long list of parameters to be considered in the study of wave-current-topography interaction to be considered. Thus, it is interesting to fix some parameters to seek for interesting regimes.

\section{Results}
\subsection{Asymptotic theory}
Asymptotic results on the interaction of a solitary waves with a topography were first reported by Grimshaw and Pelinovsky \cite{Grimshaw2002}. For the sake of completeness, we recall their main results assuming that $f(X)\rightarrow 0$ as $|X|\rightarrow\infty$. Assuming a weak force ($\epsilon\ll 1$), we seek for a slowly time-varying solitary wave with expansion
\begin{align} \label{Asymptotic}
\begin{split}
& U(x,t)=U_{0}(\xi,t)+\epsilon U_1+\cdots , \\
& \xi = x - X(t), \\
\end{split}
\end{align}
where $X(t)$ is the position of the crest of the wave. At first order, the wave profile is given by 
\begin{align} \label{Asymptotic0}
\begin{split}
& U(\xi,t)=\frac{\gamma^2}{1+B\cosh(\gamma\xi)}, \\
& \frac{dX}{dt} = c = \Delta+\gamma^2. \\
\end{split}
\end{align}
In particular,  the amplitude variation as a function of time can be obtained from the first-order momentum equation (\ref{momentum})
\begin{equation}\label{momentum0}
P_0(t)=\frac{1}{2}\int_{-\infty}^{\infty} U_{0}^{2}(\xi,t) d\xi,
\end{equation}
and its rate of change at first-order, which is given by
\begin{equation}\label{momentum0}
\frac{dP_0}{dt}=\int_{-\infty}^{\infty} U_{0}(x-X(t))\frac{df(x)}{dx} dx.
\end{equation}
Notice that $P_0$ is a function of $\gamma(t)$, thus the dynamical system (\ref{Asymptotic0})-(\ref{momentum0}) describe the amplitude and the position of the crest of the solitary wave solution. Assuming  a broad forcing %(small values of $q$ in equation (\ref{forcing}) 
the momentum equations reads
\begin{equation}\label{momentum1}
\frac{dP_0}{dt}=M_{0} \frac{df(X)}{dX} , \mbox{ where } M_{0} = \int_{-\infty}^{\infty} U_{0}(\xi,t) d\xi.
\end{equation}
Moreover, in the weak-amplitude solitary wave regime ($a\ll 1$), the quantities $M_0$, $P_0$, $\gamma$ can be obtained in explicit form
\begin{equation}\label{formulas}
M_0 = 2\sqrt{2} a^{1/2}, \;\ P_0 = \frac{2\sqrt{2}}{3} a^{3/2}, \;\ \gamma^{2} = 2 a.
\end{equation}
Therefore, the dynamical system for the amplitude and position of the crest is
\begin{align} \label{DS}
\begin{split}
& \frac{dX}{dt} = \Delta+2 a. \\
& \frac{da}{dt}=2\frac{df(X)}{dX}, 
\end{split}
\end{align}
Here, we formally consider the periodic forcing to be as the one defined in equation (\ref{forcing}). In fact, if the forcing is broad in comparison with the soliton length, the asymptotic theory is valid for any function $f(X)$ -- not only with vanishing ends. It works for periodic forcings with small values of $q$. It can be shown that the fixed points of this dynamical system are $x=\pi/2q+k\pi$, where $k$ is an integer. Centres occur aligned with the crests of the forcing while saddles aligned with the troughs of the forcing. Consequently, as we change the sign of $A$,  centres become saddles and vice-versa.
\begin{figure}[h!]
	\centering	
	\includegraphics[scale =1]{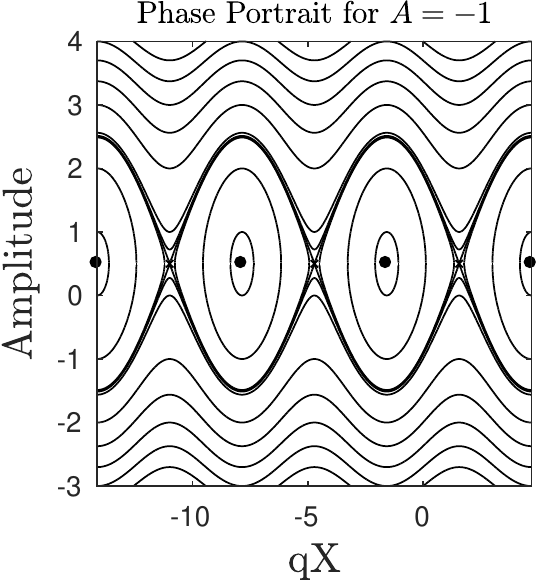}
 	\includegraphics[scale =1]{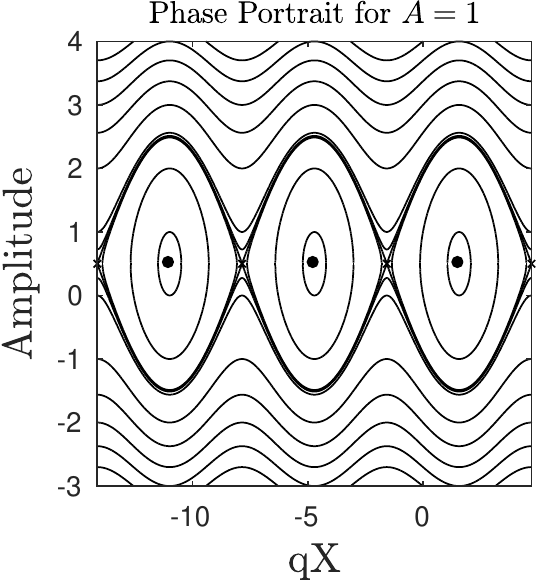}
	\caption{Phase portraits for the dynamical system (\ref{DS}) for $\Delta=-1$. Circles correspond to centres and  crosses to saddles.}
	\label{PhasePortrait}
\end{figure}

Solutions of the dynamical system (\ref{DS}) are represented by streamlines i.e., solutions are the level curves of the stream function $H(X,a)$, which is given by
\begin{equation}\label{streamfunction}
H(X,a) = -2f(X)+\Delta a+a^2.
\end{equation}
Figure \ref{PhasePortrait} displays typical phase portraits of system (\ref{DS}). Although the asymptotic results presented here are limited to the weak-amplitude case,  as it follows from Grimshaw and Pelinovsky \cite{Grimshaw2002}  qualitatively  results still hold for arbitrary amplitudes of the solitary waves and here we do not reproduce them.

\subsection{Numerical results}
Equation (\ref{UefKdV}) is solved numerically in a periodic computational domain $[-L,L]$ with a uniform grid with $N$ points using a Fourier pseudospectral method with an integrating factor \cite{Trefethen:2000}.  The computational domain is taken  large enough in order to prevent effects of the spatial periodicity. The time evolution is calculated through the Runge-Kutta fourth-order method with time step $\Delta t$. 
Typical computations are performed using $N=2^{12}$ Fourier modes $L=512$ and $\Delta t=10^{-3}$.  Solutions are compared using different number of Fourier modes ($2^{13}$ and $2^{14}$) and the results are the same. A study of the resolution of a similar numerical method can be found in \cite{Marcelo-Paul-Andre}.

We start our discussion comparing the results predicted by the asymptotic theory with the fully numerical computations. More precisely, we investigate if the centre points of the dynamical system (\ref{DS}) define trapped waves for equation (\ref{UefKdV}). Since there is a long list of parameters to be considered in the study of the interaction between a solitary wave and an external forcing, we fix a few parameters, namely, $\epsilon=0.01$, $\gamma=1$ and $q=(\pi/L)n$, where $n$ is an integer, which represents the number of waves in the interval $[-L,L]$. Notice that with these choice of parameters the initial solitary wave has amplitude $(a)$ defined in equation (\ref{amplitude}). Additionally, the initial solitary waves is chosen to be with its crest located at $x=x_0$, where $x_0$ is defined {\it a posteriori}.

\begin{figure}[h!]
	\centering	
	\includegraphics[scale =1]{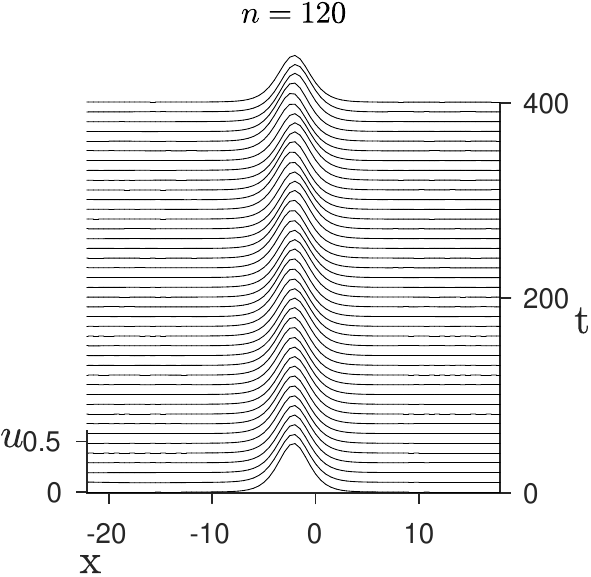}
	\includegraphics[scale =1]{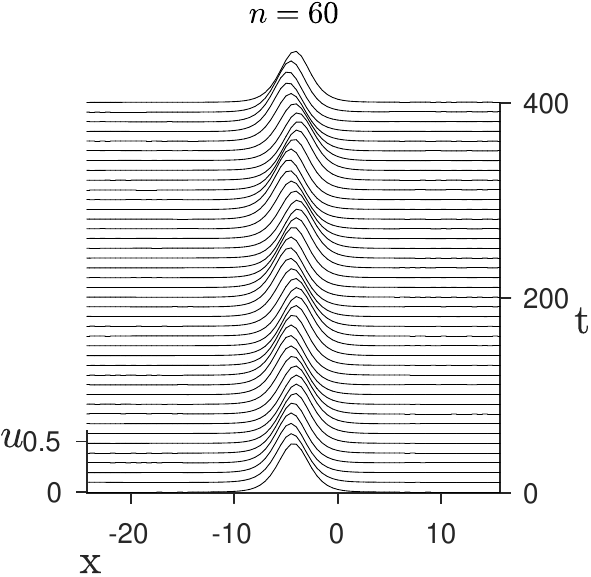}
	\includegraphics[scale =1]{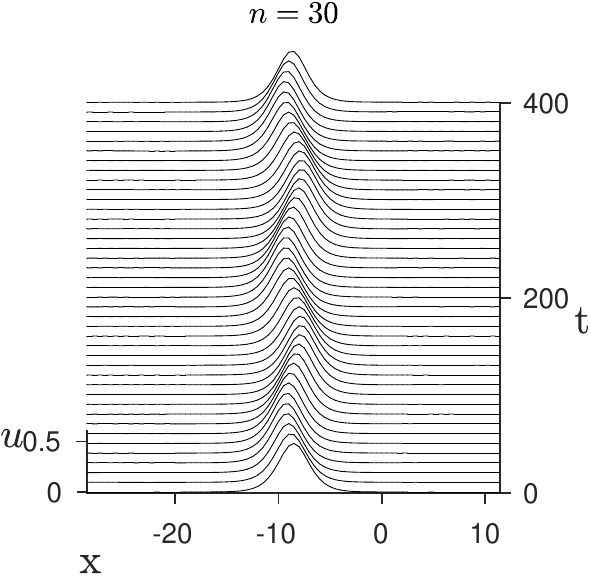}
	\includegraphics[scale =1]{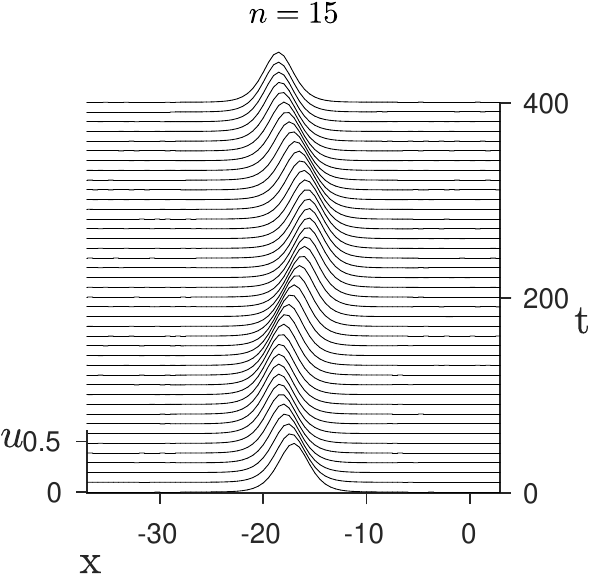}
	\caption{Solitary wave soluitons over the periodic forcing.  Parameters: $\Delta=-2a$, $x_0=-\pi/2q$, $A=-1$.}
	\label{Fig2}
\end{figure}

We initially fix the amplitude of the forcing $A=-1$.  According to the dynamical system (\ref{DS}) choosing $\Delta=-2a$ and the position of crest $x_0=-\pi/2q$ we have a centre. So, we let $q$ vary and compare the numerical results within the asymptotic framework. For large values of $n$, the solitary wave barely feels the forcing, consequently the solitary wave remains almost steady resembling the fixed point of the dynamical system (\ref{DS}). However, a growing oscillation in the position of crest indicates that this wave might move away from its initial position at large times. As we decrease the values of $n$, for instance $n=60, 30, 15$ the solitary wave bounces back and forth close to its initial position for large times with little radiation being emanated. These results are ilustrated in Figure \ref{Fig2}. It is noteworthy  that for small perturbations of $\Delta=-2a$, the solitary waves still remain trapped close to their initial positions for large times. In particular, it shows that the asymptotic theory for broad forcing  in the weak-amplitude solitary wave regime gives qualitative good results. 

\begin{figure}[h!]
	\centering	
	\includegraphics[scale =1]{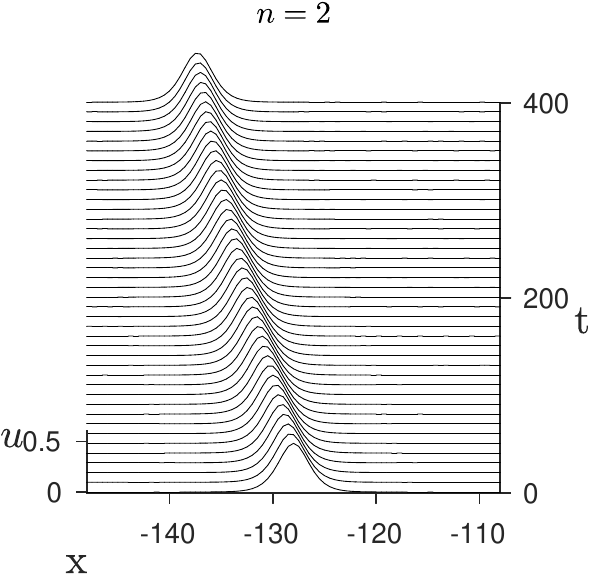}
	\includegraphics[scale =1]{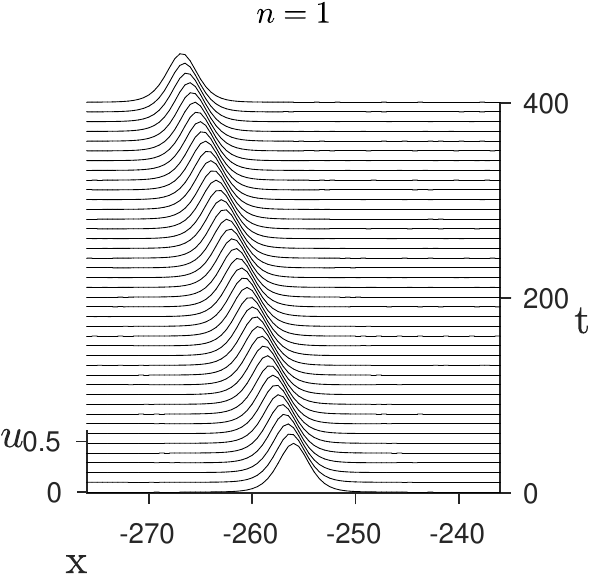}
	\caption{Solitary wave solutions over the periodic forcing.  Parameters: $\Delta=-2a$, $x_0=-\pi/2q$, $A=-1$.}
	\label{Fig2b}
\end{figure}

\begin{figure}[h!]
	\centering	
	\includegraphics[scale =1]{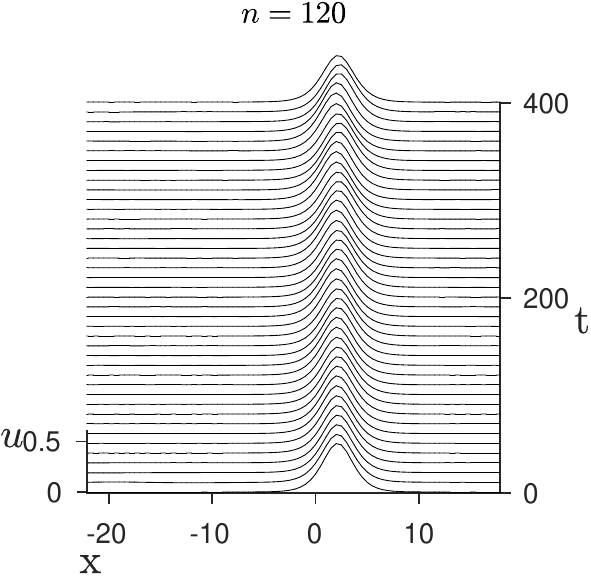}
	\includegraphics[scale =1]{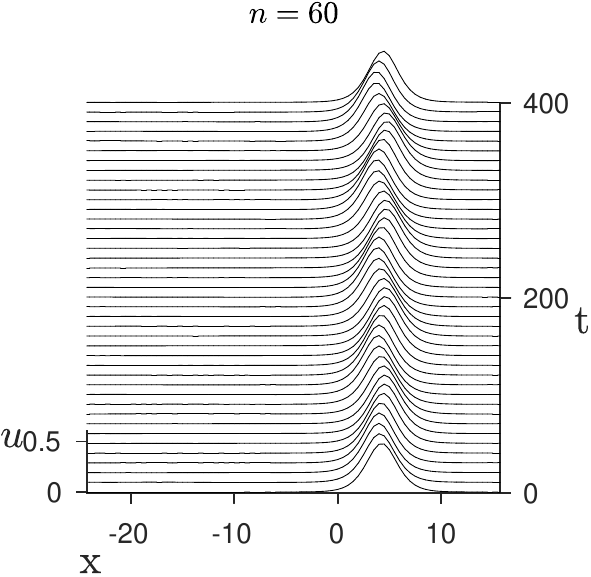}
	\includegraphics[scale =1]{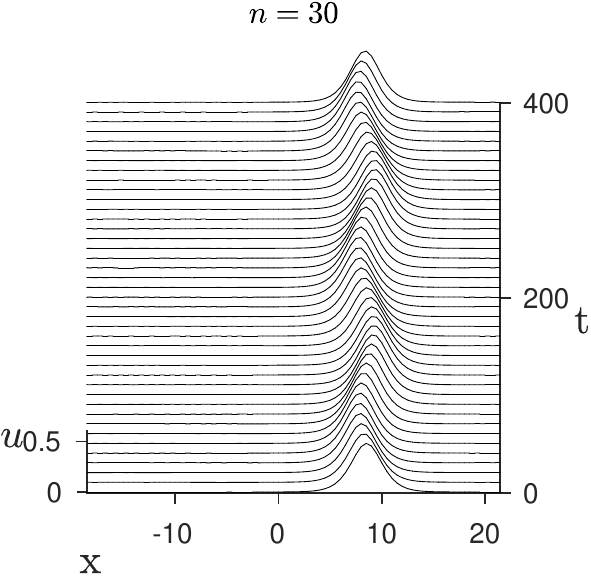}
	\includegraphics[scale =1]{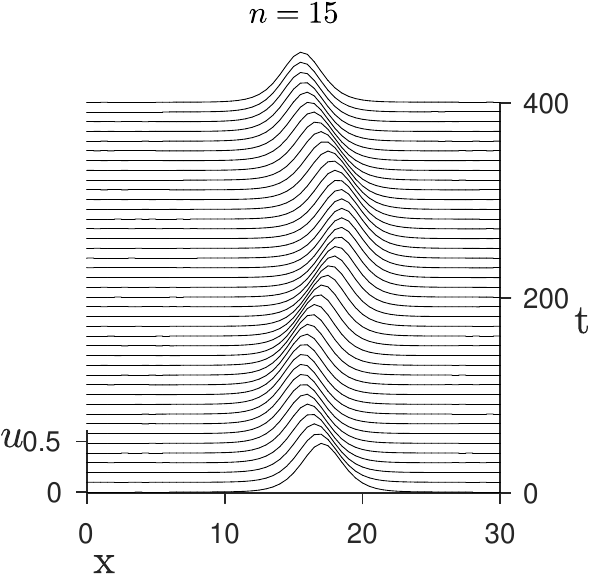}
	\includegraphics[scale =1]{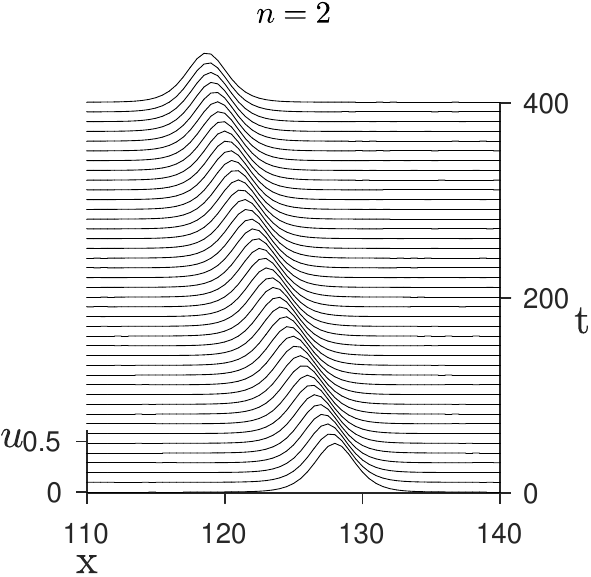}
	\includegraphics[scale =1]{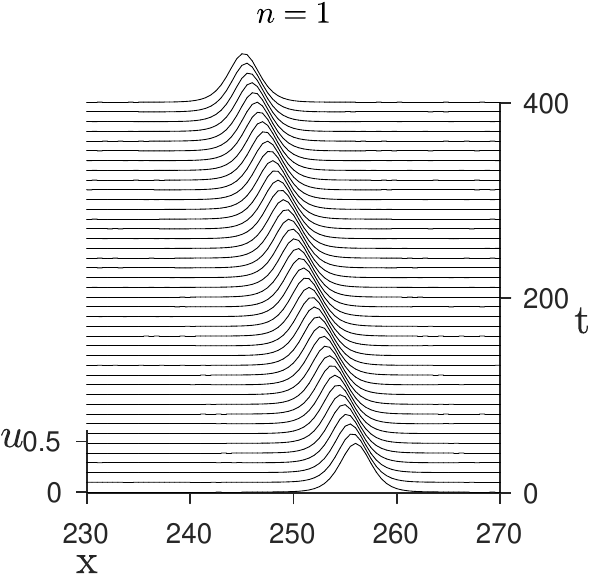}
	\caption{Solitary wave soluitons over the periodic forcing.  Parameters: $\Delta=-2a$, $x_0=\pi/2q$, $A=1$.}
	\label{Fig3}
\end{figure}

When $n$ is small, numerical results differ from the asymptotic theory. In fact,  the asymptotic method breaks for small values of $q$. It occurs because  for small values of $q$ the forcing is proportional to $\epsilon^2$. Therefore, the solitary waves is not affected by the external forcing. Figure \ref{Fig2b} displays the evolution of a solitary wave for small values of $n$. Notice that the solitary wave propagates to the left without reversing its direction. 
Similar results are observed when $A>0$ as it can be seen in Figure \ref{Fig3}.

\section{Conclusion}
In this paper, we have investigated numerically the interaction between solitary waves and a periodic forcing within the feKdV. We found that  solitary waves bounce back and forth close to their initial positions or they can move away from their initial positions. These results agree qualitatively within  the asymptotic theory.  However, when the wavenumber of the periodic forcing is too small the asymptotic fails because solitary waves can no longer be trapped within the eKdV equation.

\section{Acknowledgements}
Results described in Sect. 3a were obtained with support of RSF grant 19-12-00253.

	\section*{Declarations}
	
	\subsection*{Conflict of interest}
	The authors state that there is no conflict of interest. 
	\subsection*{Data availability}
	
	Data sharing is not applicable to this article as all parameters used in the numerical experiments are informed in this paper.

\end{document}